\documentclass[preprint,aps,nofootinbib,showpacs,amsfonts,epsf]{revtex4}

\input{epsf.tex}

\newcommand{\E}{{\cal{E}}}
\newcommand{\s}{\sigma}

\newcommand{\be}{\begin{equation}}
\newcommand{\ee}{\end{equation}}
\newcommand{\bea}{\begin{eqnarray}}
\newcommand{\eea}{\end{eqnarray}}
\newcommand{\ba}{\begin{array}}
\newcommand{\ea}{\end{array}}
\def\J#1#2#3#4{{#1} {\bf #2}, #3 (#4)}
\def\PRD{Phys. Rev. D}
\def\PR{Phys. Rev.}
\def\PRL{Phys. Rev. Lett.}

\def\JMP{J. Math. Phys.}

\def\MZ{Math. Z.}

\def\CQG{Class. Quantum Grav.}

\def\NPB{Nucl. Phys. B}
\def\MZ{Math. Zeits.}

\def\JHEP{J. High Energy Phys.}

\begin{document}
\draft
\title{Thermodynamics of two aligned Kerr-Newman black holes}

\author{H. Garc\'ia-Compe\'an, V.~S.~Manko and C. J. Ram\'irez-Valdez}
\address{$^\dagger$Departamento de F\'\i sica, Centro de Investigaci\'on y
de Estudios Avanzados del IPN, A.P. 14-740, 07000 Ciudad de
M\'exico, Mexico}

\begin{abstract}
Using the notion of thermodynamic length, the first law of
thermodynamics is consistently derived for two binary
configurations of equal Kerr-Newman black holes separated by a
massless strut. Like in the electrostatic systems of two
Reissner-Nordstr\"om black holes and stationary vacuum systems of
two Kerr black holes considered earlier, the thermodynamic length
$\ell$ turns out to be defined by the same simple formula
$\ell=L\exp(\gamma_0)$, $L$ being the coordinate length of the
strut and $\gamma_0$ the value of the metric function $\gamma$ on
the strut, which permits the elaboration of $\ell$ in a concise
analytic form. The expression of the free energy in the
case of two generic Kerr-Newman black holes is also proposed.
\end{abstract}


\maketitle

\section{Introduction}

In our previous paper \cite{RGM} we have performed analytical
derivation of the first law of thermodynamics for the binary
configurations of Kerr black holes \cite{Ker} with the aid of the
notion of thermodynamic length introduced by Appels, Gregory and
Kubiz\v{n}\'ak \cite{AGK}, and we have shown that in the
stationary vacuum case the thermodynamic length is defined by the
same formula as in the electrostatic case analyzed in \cite{KZe}.
It was also suggested in \cite{RGM} that thermodynamics of the
Kerr-Newman (KN) black holes \cite{New} in binary systems could
probably be treated as readily as in the aforementioned two cases,
and the aim of the present paper is to demonstrate that this is
indeed true when the usual approach to the thermodynamic length is
employed in the derivation procedures.

The binary configurations of our interest are obtainable from the
double-Kerr-Newman solution \cite{MMR} by imposing certain
conditions on the metric functions. Since the latter conditions
have not yet been solved analytically in the general case of two
black-hole constituents separated by a strut \cite{Isr}, our consideration will be
restricted to two different configurations of equal
counterrotating KN black holes, in one of which the charges of the
constituents have the same signs and in the other the charges have
opposite signs. The first configuration is described by the
Bret\'on-Manko (BM) electrovac solution \cite{BMa} for which a
physical parametrization was worked out in \cite{MRR}; on the
other hand, the binary system with opposite charges is described
by a 4-parameter specialization of the Ernst-Manko-Ruiz
equatorially antisymmetric solution \cite{EMR} whose physical
parametrization was worked out in \cite{MRS}. Actually, both the
above electrovac solutions may be considered as two different
charging generalizations of the vacuum BM spacetime studied in our
previous paper \cite{RGM}. Mention also that one of the virtues of
the solutions for equal counterrotating KN black holes is that the rotation
parameter entering these solutions represents exactly the angular
momentum per unit mass of each black-hole constituent, which
simplifies considerably the related symbolic computations.

Our paper is organized as follows. In Sec.~II we derive the first
law of thermodynamics for the binary system of KN black holes
described by the BM solution. Here we also give the form of the
Euclidean action for this system and show that the corresponding
free energy involving explicitly the thermodynamic length is able
to reproduce {\it exactly} the expressions of the entropy and
interaction force. In Sec.~III the analogous thermodynamic
analysis is performed for the binary configuration of
counterrotating KN black holes endowed with opposite charges, and
it confirms the main conclusions of the previous section. The
discussion of the results obtained, with possible extensions to
the binary systems of unequal KN black holes can be found in Sec.~IV.
Concluding remarks are given in Sec.~V.

\section{Two equal counterrotating KN black holes carrying the
same charges}

The BM 4-parameter electrovac solution for two identical
counterrotating KN black holes is described by the line element
\be d s^2=f^{-1}[e^{2\gamma}(d\rho^2+d z^2)+\rho^2 d\varphi^2]-f(d
t-\omega d\varphi)^2, \label{Papa} \ee
with the metric functions $f$, $\gamma$ and $\omega$ of the form
\cite{MRR}
\bea f&=&\frac{A\bar A-B\bar B+C\bar C}{(A+B)(\bar A+\bar B)},
\quad e^{2\gamma}=\frac{A\bar A-B\bar B+C\bar
C}{16R^4\sigma^4R_+R_-r_+r_-}, \quad \omega=-\frac{{\rm
Im}[2G(\bar A+\bar B)+C\bar I]}{A\bar A-B\bar B+C\bar C},
\nonumber\\
A&=&(M^2-Q^2)[4\sigma^2(R_+R_-+r_+r_-)+R^2(R_+r_++R_-r_-)]
+[\sigma^2(R^2-4M^2+4Q^2)  \nonumber\\
&&-J^2R^2\mu](R_+r_-+R_-r_+)
-2iJR\sigma\mu(MR+2M^2-Q^2)(R_+r_--R_-r_+), \nonumber\\
B&=&2MR\sigma[R\sigma (R_++R_-+r_++r_-)-\lambda(R_+-R_--r_++r_-)],
\nonumber\\ C&=&QB/M, \nonumber\\
G&=&-zB+R\sigma\{(2M^2-Q^2)[2\sigma(r_+r_--R_+R_-)+R(R_-r_--R_+r_+)]
\nonumber\\
&&+M(R+2\sigma)(R\sigma-\lambda)(R_+-r_-)
+M(R-2\sigma)(R\sigma+\lambda)(R_--r_+)\},
\nonumber\\
I&=&\frac{Q}{M}\{G+RQ^2\sigma[2\sigma(r_+r_--R_+R_-)+R(R_-r_--R_+r_+)]\},
\nonumber\\ \lambda&\equiv&2(M^2-Q^2)+iJ\mu(MR+2M^2-Q^2), \nonumber\\
R_\pm&=&\sqrt{\rho^2+\left(z+\frac{1}{2}R\pm\sigma\right)^2},
\quad r_\pm=\sqrt{\rho^2+\left(z-\frac{1}{2}R\pm\sigma\right)^2},
\label{metric1} \eea
where the arbitrary real parameters $M$ and $Q$ are, respectively,
the mass and electric charge of each black hole, while $J$ is the
angular momentum of the lower black hole (the angular momentum of
the upper black hole is $-J$); $R$ denotes the coordinate distance
between the centers of black holes (see Fig.~1). The constant
quantity $\s$ determining horizon's length is defined by the
expression
\be \sigma=\sqrt{M^2-Q^2-J^2\mu}, \quad
\mu\equiv\frac{R^2-4M^2+4Q^2}{(MR+2M^2-Q^2)^2}. \label{sigma1} \ee
Besides, the $t$ and $\varphi$ components of the electromagnetic
4-potential are given by the formulas
\be A_t=-{\rm Re}\left(\frac{C}{A+B}\right), \quad A_\varphi={\rm
Im}\left(\frac{I}{A+B}\right). \label{Atf} \ee

Since we are interested exclusively in the black-hole sector of
the BM solution, then $\s$ must be a real quantity, which means
that the parameters of the solution are subject to the constraint
$\s^2>0$. Note that in the above formulas we have introduced
explicitly the angular momentum $J$ (of the lower black hole)
instead of the parameter $a$ used in \cite{MRS}, which is more
suitable for the analysis of thermodynamics of the BM solution. In
the absence of electric charge ($Q=0$), one recovers from
(\ref{Papa})-(\ref{sigma1}) the vacuum solution for two
counterrotating Kerr black holes considered in our previous paper,
and in Appendix we give its representation obtainable from the
general formulas for binary configurations of nonequal black holes
that might represent interest for applications.

Each black hole of the BM solution satisfies Smarr's mass formula
\cite{Sma}
\be M=2TS+2\Omega J+\Phi_e Q, \label{Sma} \ee
where $T$ is the temperature and $S$ the entropy of each black
hole, $\Omega$ is horizon's angular velocity of the lower black
hole ($-\Omega$ of the upper black hole) and $\Phi_e$ is the
electric potential. For the quantities $S$, $\Omega$ and $\Phi_e$,
as well as for the interaction force ${\cal F}$, the paper
\cite{MRR} gives the following expressions:
\bea T&=&\frac{R\s}{2\pi\Delta}, \quad S=\frac{\pi\Delta}{R},
\quad \Omega=\frac{J\mu(MR+2M^2-Q^2)}{\Delta}, \nonumber\\
\Phi_e&=&\frac{Q[(R+2M)(M+\s)-2Q^2]}{\Delta}, \quad {\cal F}=
\frac{M^2-Q^2}{R^2-4M^2+4Q^2}, \nonumber\\
\Delta&\equiv&2M(R+2M)(M+\s)-Q^2(R+4M+2\s). \label{TS1} \eea
To derive consistently the first law of thermodynamics and find
the explicit form of the corresponding thermodynamic length $\ell$
for this binary configuration, we must take differentials of the
four quantities $S$, $\Omega$, $\Phi_e$ and ${\cal F}$ by
considering these as functions of $M$, $J$, $R$ and $Q$, and solve
the resulting system for $dM$, $d\Omega$, $d\Phi_e$ and $dR$, of
which for the differential $dM$ we get the expression
\be dM=TdS+\Omega dJ+\Phi_e dQ-\frac{1}{2R^2}
(R-2\s)(R^2-4M^2+4Q^2)d{\cal F}. \label{dM1} \ee
Taking into account the equality of black holes, we finally arrive
at the desired first law and thermodynamic length in the form
\bea dM_T&=&2TdS+2\Omega dJ+2\Phi_e dQ-\ell d{\cal F}, \nonumber\\
M_T&=&2M, \quad \ell=(R-2\s)\left(1-\frac{4(M^2-Q^2)}{R^2}\right),
\label{L1} \eea
and it is not difficult to show using formulas
(\ref{metric1})-(\ref{sigma1}) that, similar to the electrostatic
case considered in \cite{KZe} or the pure vacuum case considered
in \cite{RGM}, the thermodynamic length $\ell$ is defined by the
formula
\be \ell=Le^{\gamma_0}, \label{el1} \ee
where $L=R-2\s$ is the coordinate length of the strut (see
Fig.~1), and $\gamma_0$ is the value of the metric function
$\gamma$ on the strut. It might be tempting to think looking at (\ref{L1}) that
$\ell=L$ when $M^2=Q^2$; however, in this particular case the
conical singularity disappears (since ${\cal F}=0$) and the two
constituents are in neutral equilibrium independent of the
separation distance due to balance of the gravitational and
electromagnetic forces: for nonvanishing angular momentum the
resulting equilibrium configuration is described by the
Parker-Ruffini-Wilkins metric \cite{PRW} for hyperextreme KN
sources, while for vanishing angular momentum the equilibrium is
between two extreme charged black holes \cite{Maj,Pap}.

To gain a better insight into thermodynamics of the BM binary
configuration, we have computed, via standard procedures, the
Euclidean action \cite{GHa} corresponding to this configuration,
yielding the simple expression
\be I=\beta(M-\Phi_e Q+\ell{\cal F}), \label{I1} \ee
where $\beta=1/T$, and the quantities $\Phi_e$, ${\cal F}$ and
$\ell$ are defined by the formulas (\ref{TS1}) and (\ref{dM1}).
Hence, taking into account that $I=\beta W$, $W$ being the free
energy, we get
\be W=M-\Phi_e Q+\ell{\cal F}. \label{W1} \ee

To find the variation of the thermodynamic potential $W$, we must first take
differentials of the quantities $\ell$, $T$, $\Omega$ and $\Phi_e$
by considering these as functions of the parameters $M$, $J$, $Q$ and $R$ and
then express $dM$, $dJ$, $dQ$ and $dR$ in terms of the
differentials $d\ell$, $dT$, $d\Omega$ and $d\Phi_e$ by solving the
linear system of four algebraic equations. Expressing subsequently the
differential $dW$ in terms of the latter four differentials, we
finally arrive at the formula
\be dW=-2SdT-2Jd\Omega-2Qd\Phi_e+{\cal F}d\ell, \label{dW1} \ee
whence it follows that
\be S=-\left.\frac{1}{2}\frac{\partial W}{\partial T} \right|_{\Omega,\Phi_e,\ell},
\quad J=-\left.\frac{1}{2}\frac{\partial W}{\partial\Omega}
\right|_{T,\Phi_e,\ell}, \quad Q=-\left.\frac{1}{2}\frac{\partial W}{\partial\Phi_e}
\right|_{T,\Omega,\ell}, \quad {\cal F}=\left.\frac{\partial W}{\partial\ell}
\right|_{T,\Omega,\Phi_e}. \label{SJQ1} \ee
It is remarkable that the relations (\ref{dW1}) and (\ref{SJQ1})
are verified {\it exactly} due to the use in (\ref{W1}) of the
thermodynamic length $\ell$. We also note that derivation of the
formula (\ref{dW1}) is computationally a more complicated task than obtaining
the first law of thermodynamics in the form (\ref{L1}).

\section{Two equal counterrotating KN black holes with
opposite charges}

While the BM solution dates back to 1995, the binary configuration
of equal KN counterrotating black holes with opposite electric
charges was identified and correctly described only quite recently
\cite{MRS}. It is determined by the metric functions of the form
\bea f&=&\frac{A\bar A-B\bar B+C\bar C}{(A+B)(\bar A+\bar B)},
\quad e^{2\gamma}=\frac{A\bar A-B\bar B+C\bar C}
{16R^4\s^4R_+R_-r_+r_-}, \quad \omega=-\frac{{\rm Im}[2G(\bar
A+\bar B)+C\bar I]} {A\bar A-B\bar B+C\bar C}, \nonumber \\
A&=&R^2(M^2-Q^2\nu)(R_+-R_-)(r_+-r_-)+
4\s^2(M^2+Q^2\nu)(R_+-r_+)(R_--r_-) \nonumber\\
&&+2R\s[R\s(R_+r_-+R_-r_+)+iJ\mu(R_+r_--R_-r_+)], \nonumber\\
B&=&2MR\s[R\s(R_++R_-+r_++r_-)-(2M^2-iJ\mu)(R_+-R_--r_++r_-)],
\nonumber\\
C&=&2C_0R\s[(R+2\s)(R\s-2M^2-iJ\mu)(r_+-R_-) +(R-2\s) \nonumber\\
&&\times(R\s+2M^2+iJ\mu)(r_--R_+)], \nonumber\\
G&=&-zB+R\s\{R(2M^2-Q^2\nu)(R_-r_--R_+r_+)
+2\s(2M^2+Q^2\nu)(r_+r_--R_+R_-)
\nonumber\\
&&+M[(R+2\s)(R\s-2M^2+iJ\mu)+2(R-2\s)Q^2\nu](R_+-r_-)
\nonumber\\
&&+M[(R-2\s)(R\s+2M^2-iJ\mu)-2(R+2\s)Q^2\nu](R_--r_+)\},
\nonumber\\
I&=&-zC+2C_0M[R^2(2M^2-2\s^2+iJ\mu)(R_+r_++R_-r_-)
\nonumber\\
&&+2\s^2(R^2-4M^2-2iJ\mu)(R_+R_-+r_+r_-)]-C_0(R^2-4\s^2) \nonumber\\
&&\times\{2M[R\s(R_+r_--R_-r_+)+(2M^2+iJ\mu)(R_+r_-+R_-r_+)] +R\s[R\s \nonumber\\
&&\times(R_++R_-+r_++r_-)+(6M^2+iJ\mu)(R_+-R_--r_++r_-) +8MR\s]\},
\nonumber\\
R_\pm&=&\sqrt{\rho^2+\left(z+\frac{1}{2}R\pm\sigma\right)^2},
\quad r_\pm=\sqrt{\rho^2+\left(z-\frac{1}{2}R\pm\sigma\right)^2},
\label{metric2} \eea
where the dimensionless quantities $\mu$, $\nu$ and $C_0$ are
defined as
\be \mu=\frac{R^2-4M^2}{M(R+2M)+Q^2}, \quad
\nu=\frac{R^2-4M^2}{(R+2M)^2+4Q^2}, \quad
C_0=-\frac{Q(R^2-4M^2+2iJ\mu)}{(R+2M)(R^2-4\s^2)}, \label{mnc} \ee
and the horizons' half length $\s$ is given by the expression
\be \s=\sqrt{M^2-\left(\frac{J^2[(R+2M)^2+4Q^2]}
{[M(R+2M)+Q^2]^2}+Q^2\right)\frac{R-2M}{R+2M}}. \label{sigma2} \ee
The arbitrary parameters of the solution are $M$, $J$, $Q$ and $R$
which have the same meaning as in the previous section, with the
only reservation that now $J$ is the angular momentum of the upper
constituent ($-J$ of the lower constituent, whose electric charge
is $-Q$, see Fig.~2). Note that in the above formulas $J$ has been
introduced explicitly as the rotation parameter instead of the
parameter $a$ used in \cite{MRS}. The electromagnetic field is defined by
Eqs.~(\ref{Atf}) of the previous section. The black-hole sector of the
solution (\ref{metric2})-(\ref{sigma2}), as usual, corresponds to
$\s^2>0$. One can see that the formulas describing the case of
opposite charges are somewhat more complicated than those of the
case of identical charges.

The two KN black holes have the same temperature $T$ and entropy
$S$, but their angular momenta, charges, horizon's angular
velocities and electric potentials have opposite signs;
nonetheless, the Smarr formula (\ref{Sma}) is verified by each
black hole, and below we give the form of $T$, $S$, $\Omega$,
$\Phi_e$ obtained in \cite{MRS} for the upper constituent, to
which the expression of the interaction force ${\cal F}$ is also added:
\bea T&=&\frac{R\s[(R+2M)^2+4Q^2]}
{2\pi(R+2M)^2[2(M+\s)(MR+2M^2+Q^2)-Q^2(R-2M)]},  \nonumber\\
S&=&\frac{\pi}{R(R+2\s)}\left((R+2M)^2(M+\s)^2
+\frac{J^2(R^2-4M^2)^2}{(MR+2M^2+Q^2)^2}\right),  \nonumber\\
\Omega&=&\frac{J[2(M-\s)(MR+2M^2+Q^2)-Q^2(R-2M)]}
{(4J^2+Q^4)(MR+2M^2+Q^2)},  \nonumber\\
\Phi_e&=&\frac{Q[Q^2(M-\s)(MR+2M^2+Q^2)+2J^2(R-2M)]}
{(4J^2+Q^4)(MR+2M^2+Q^2)}, \nonumber\\
{\cal F}&=&\frac{M^2(R+2M)^2+Q^2R^2}
{(R+2M)^2(R^2-4M^2)}.   \label{TS2} \eea
The procedure of obtaining the first law of thermodynamics
for this second binary configuration involving the quantities $S$,
$\Omega$, $\Phi_e$ and ${\cal F}$ is fully analogous to the
already considered in the previous section during the derivation
of equations (\ref{L1}), so we will restrict ourselves to just writing
down the final result for the first law and corresponding
thermodynamic length:
\bea dM_T&=&2TdS+2\Omega dJ+2\Phi_e dQ-\ell d{\cal F}, \nonumber\\
M_T&=&2M, \quad \ell=\frac{(R-2\s)(R+2M)^2(R^2-4M^2)}
{R^2[(R+2M)^2+4Q^2]}.
\label{L2} \eea
A direct check shows that the above $\ell$ verifies formula
(\ref{el1}).

A thorough straightforward analysis of the Euclidean action
and free energy of the solution reveals that these have exactly
the same structure defined by formulas (\ref{I1}) and (\ref{W1})
as in the case of the BM configuration, including the formula
(\ref{dW1}) for the variation of $W$. Therefore, we have got
another evidence that the thermodynamic length determines in a
natural and consistent way the thermodynamical properties of the
binary systems of KN black holes.

\section{Discussion}

While the case of equal black holes is very suitable for the
thermodynamic analysis because the binary configurations are in
thermal equilibrium allowing for the introduction of the free
energy potential through the Euclidean action, the question still
arises of how such a potential could be defined when a binary
configuration is composed of unequal black holes having different
temperatures. We recall in this respect that even in the simplest
case of the double-Schwarzschild solution \cite{BWe}
the paper \cite{CPe} for instance does not analyze the general case of two
Schwarzschild black holes, most probably because of the
difficulties in the description of the Euclidean action when a
binary system is not in thermal equilibrium. Nonetheless it is
remarkable that, thanks to the thermodynamic length, the free
energy $W$ in the case of nonequal black-hole constituents can be
easily guessed from the expressions of $W$ obtained for various
binary configurations of equal constituents. Indeed, in one
assumes that the contribution of the strut in $W$ is defined by
the term $\ell{\cal F}$, then the expression for $W$ in the general
double-Schwarzschild solution takes the form ($m_i$ are the masses
of black holes and $R$ the distance between the black hole
centers)
\bea W&=&\frac{1}{2}(m_1+m_2)+\ell{\cal F}, \nonumber\\
\ell&=&(R-m_1-m_2)\frac{R^2-(m_1+m_2)^2}
{R^2-(m_1-m_2)^2}, \quad {\cal F}=\frac{m_1m_2}
{R^2-(m_1+m_2)^2},
\label{W3} \eea
where the above expression for $\ell$ was found in \cite{KZe}.
Then, using formulas for the temperatures $T_1$ and $T_2$ of the
black holes from \cite{KZe} and the expression of $\ell$ from
(\ref{W3}) we shall get, via the procedure described in Sec.~II,
that
\bea dW&=&-S_1dT_1-S_2dT_2+{\cal F}d\ell, \nonumber\\
S_1&=&\frac{4\pi m_1(R+m_1+m_2)}{R+m_1-m_2}, \quad
S_2=\frac{4\pi m_2(R+m_1+m_2)}{R-m_1+m_2}, \label{dW3} \eea
$S_1$ and $S_2$ being the entropies of the Schwarzschild black
holes, so that
\be S_1=-\left.\frac{\partial W}{\partial T_1} \right|_{T_2,\ell},
\quad S_2=-\left.\frac{\partial W}{\partial T_2} \right|_{T_1,\ell}. \label{S12} \ee

Note that for the derivation of (\ref{dW3}) it is not sufficient
only to calculate correctly the contribution of the strut in
(\ref{W3}), namely,
\be \frac{m_1m_2(R-m_1-m_2)}{R^2-(m_1-m_2)^2}, \label{cs} \ee
but it is of paramount importance to present the quantity
(\ref{cs}) in the form $\ell{\cal F}$, because any other
presentation of (\ref{cs}), say, in the form $LF$, with
$L=R-m_1-m_2$ and $F=m_1m_2/[R^2-(m_1-m_2)^2]$, will not permit to
derive the correct expression for $dW$ leading to (\ref{S12})
exactly (even in the case of equal constituents).

Moreover, we have also investigated the issue of the free energy
potential for all the binary configurations of Kerr black holes
considered in our previous paper \cite{RGM}, as well as for the
double-Reissner-Nordstr\"om solution \cite{BMA,Man} analyzed in
\cite{KZe}, and this endeavor has made us tentatively conclude that the
expression of the free energy in the binary system of generic KN
black holes separated by a massless strut and described by seven
independent parameters $M_i$, $J_i$, $Q_i$ and R, $i=1,2$, must
have the form
\be W=\frac{1}{2}\sum_{i=1}^2(M_i-Q_i\Phi_i)+\ell{\cal F},
\label{W4} \ee
while the variation of $W$ is given by the formula
\be dW=-\sum_{i=1}^2(S_idT_i+J_id\Omega_i
+Q_id\Phi_i)+{\cal F}d\ell,
\label{dW4} \ee
where $S_i$, $T_i$, $\Omega_i$ and $\Phi_i$ stand, respectively,
for the individual entropies, temperatures, horizon's angular
velocities and electric potentials of the black holes (of course,
$M_i$, $J_i$ and $Q_i$ are the masses, angular momenta and
electric charges of black holes). Besides, the corresponding first
law of thermodynamics reads as
\be d(M_1+M_2)=\sum_{i=1}^2(T_idS_i +\Omega_i dJ_i+\Phi_i dQ_i)
-\ell d{\cal F}. \label{dM4} \ee
The thermodynamic length $\ell$ entering (\ref{W4})-(\ref{dM4}) is
universally defined by the formula (\ref{el1}) first obtained in
\cite{KZe}, and its geometrical interpretation proposed in the
latter paper is the area of the worldsheet of the strut per unit time.

\section*{APPENDIX}

Here we give a possible concise representation of the vacuum BM
solution worked out with the aid of the general formulas of the
paper \cite{RGM}. Its Ernst potential \cite{Ern} is determined by the
formulas
\bea \E&=&(A-B)/(A+B), \nonumber\\
A&=&R^2(R_+-R_-)(r_+-r_-)
-4\s^2(R_+-r_+)(R_--r_-), \nonumber\\
B&=&2R\s[(R+2\s)(R_--r_+)-(R-2\s)(R_+-r_-)],  \nonumber\\
r_\pm&=&\frac{\pm\s-M+ia\mu} {\pm\s-M-ia\mu}\,\tilde r_\pm, \quad
R_\pm=-\frac{\pm\s+M-ia\mu} {\pm\s+M+ia\mu}\,\tilde R_\pm, \nonumber\\
\tilde r_\pm&=&\sqrt{\rho^2+ \left(z-\frac{1}{2}R\pm\s\right)^2},
\quad \tilde R_\pm=\sqrt{\rho^2+
\left(z+\frac{1}{2}R\pm\s\right)^2}, \label{Eps} \eea
where $M$ is the mass of each Kerr constituent, $a$ the angular
momentum per unit mass of the lower constituent and $R$ the
coordinate distance between the centers of the sources; the
horizon's half length has the form
\be \sigma=\sqrt{M^2-a^2\mu}, \quad \mu\equiv\frac{R-2M}{R+2M},
\label{sigma} \ee
while the corresponding metric functions are defined by the
expressions
\bea f&=&\frac{A\bar A-B\bar B}{(A+B)(\bar A+\bar B)}, \quad
e^{2\gamma}=\frac{A\bar A-B\bar B}{K_0\tilde R_+\tilde R_-\tilde
r_+\tilde r_-}, \quad \omega=-\frac{2{\rm Im}[G(\bar A+\bar
B)]}{A\bar A-B\bar B}, \nonumber\\ G&=&-zB +R\s[2R(R_-r_--R_+r_+)
+4\s(R_+R_--r_+r_-)\nonumber\\
&&-(R^2-4\s^2)(R_+-R_--r_++r_-)], \nonumber\\
K_0&=&16\s^4R^4/M^4. \label{metric} \eea

This solution represents the pure vacuum limit of both electrovac
metrics considered in our paper.

\section*{Acknowledgments}

This work was partially supported by Project~128761 from CONACyT
of Mexico.

\begin{figure}[htb]
\centerline{\epsfysize=75mm\epsffile{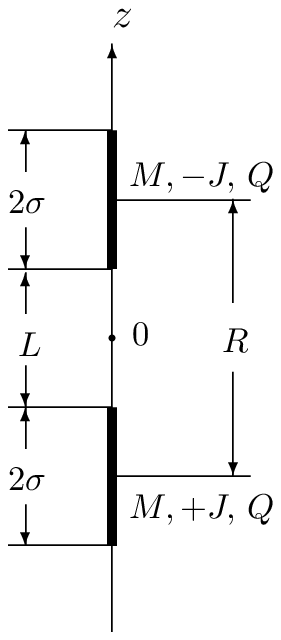}} \caption{Location
of two equal counterrotating KN black holes with the same charges on
the symmetry axis.}
\end{figure}

\begin{figure}[htb]
\centerline{\epsfysize=75mm\epsffile{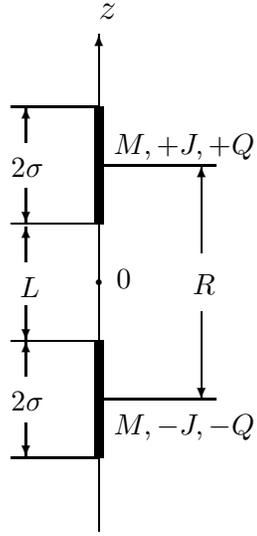}} \caption{Location
of two equal counterrotating KN black holes with opposite
charges on the symmetry axis.}
\end{figure}

\end{document}